\documentclass[10pt, a4paper, twocolumn]{article}

\usepackage[utf8]{inputenc}
\usepackage[T1]{fontenc}
\usepackage[english]{babel}
\usepackage{lmodern}
\usepackage[margin=1.8cm]{geometry}
\usepackage{amsmath, amssymb, amsthm}
\usepackage{mathtools}
\usepackage{booktabs}
\usepackage{graphicx}
\usepackage{mathrsfs}
\usepackage{siunitx} 
\usepackage{soul}
\usepackage{float}
\usepackage{stfloats}

\usepackage{tikz}
\usetikzlibrary{angles, quotes, arrows.meta}

\usepackage{microtype} 
\usepackage{caption}
\usepackage{csquotes}

\usepackage{xspace}
\newcommand{\HDirac}{{\sc HDirac}\xspace}
\newcommand{\Affect}{{\sc Aff3ct}\xspace}
\newcommand{\Eigen}{{\sc Eigen}\xspace}
\newcommand{\Dalek}{{\sc Dalek}\xspace}
\newcommand{\StreamPU}{{\sc StreamPU}\xspace}

\usepackage[
    style=authoryear,
    sorting=nyt,
    giveninits=true,
    terseinits=true,
    maxbibnames=99, 
    uniquename=false,
    uniquelist=false
]{biblatex}

\renewbibmacro{in:}{}

\DeclareFieldFormat{biblabeldate}{#1}

\let\oldcite\cite
\renewcommand{\cite}[1]{(\oldcite{#1})}
\newcommand{\R}{\mathcal{R}}

\title{\vspace{-1cm} 
Multidimensional Reconciliation in Continuous-Variable QKD: Review, Coding Schemes, and Open-Source Simulation
}

\usepackage{authblk}

\author[1,2]{Lucien Martial}
\author[2]{Alexis Rosio}
\author[2]{Eleni Diamanti}
\author[2]{Adrien Cassagne}
\author[1]{Baptiste Gouraud}

\affil[1]{Exail, F-25000, Besançon, France}
\affil[2]{Sorbonne Université, CNRS, LIP6, F-75005 Paris, France}

\date{}

\begin{document}

\maketitle

\begin{abstract}
    Continuous-variable quantum key distribution (CV-QKD) requires highly efficient reconciliation techniques to operate at low signal-to-noise ratios and long distances. Multidimensional reconciliation addresses this challenge by transforming the physical Gaussian quantum channel into a virtual binary-input additive white Gaussian noise (BIAWGN) channel, enabling the use of modern error-correcting codes.
    In this work, we review the principles of multidimensional reconciliation, with a particular focus on high-dimensional constructions beyond the algebraic dimensions {1, 2, 4, 8}. We describe the construction of the virtual channel, discuss practical coding schemes for reverse reconciliation, and analyse their integration with linear error-correcting codes. We also present an open-source simulation framework, \HDirac, implementing multidimensional reconciliation for arbitrary dimensions, and use it to evaluate state-of-the-art LDPC codes. The results highlight key trade-offs between dimension, reconciliation efficiency, and frame error rate, providing practical guidance for CV-QKD system design.

\end{abstract}

\section{Introduction}

Classical cryptography typically relies on hard computational problems and on the assumption that an adversary (Eve) has limited computational power and time. As a result, its long-term security cannot be ensured, given the steady progress of classical computing and the advent of quantum computing.
In contrast, in the paradigm of information-theoretic security (ITS), the information accessible to Eve is bounded, even if Eve has unlimited computational power. A central example of an ITS cryptographic primitive is quantum key distribution (QKD), which enables two distant parties to establish a shared secret key based on the laws of quantum physics.

QKD based on the encoding of information in continuous variables (CV-QKD) offers specific advantages over QKD based on encoding in single-photon properties, known as discrete variable quantum key distribution (DV-QKD). More specifically, in CV-QKD protocols, information is encoded in the quadratures of the electromagnetic field, usually via the modulation of the coherent states emitted by a laser. CV-QKD systems therefore use standard optical telecommunications components and techniques, benefiting from their low cost and high reliability, enabling high secret key rates and, in particular, eliminating the need for single-photon detectors used in DV-QKD.
However, these advantages come with an increased complexity in the post-processing stages, particularly in the high-loss (long-distance) regime. While in DV-QKD data can be post-selected on photon detection events, in CV-QKD it is necessary to manage signals largely dominated by quantum vacuum noise. CV-QKD thus requires more sophisticated reconciliation, the error correction step which is essential to ensure that both parties ultimately share a key that is strictly identical.

Multidimensional reconciliation has become the method of choice for long-distance CV-QKD. It was proposed in~\cite{Leverrier2008} as an alternative to slice reconciliation~\cite{Assche2004}, better suited in the large-SNR/short-distance regime. While slice reconciliation attempts to recover multiple bits per channel use, multidimensional reconciliation transforms the physical quantum channel into a virtual channel that approximates a binary input additive white Gaussian noise (BIAWGN) channel. Reconciliation consequently benefits from the extended research conducted on error-correction for classical communications on a BIAWGN channel. The core principle involves mapping a subvector of size $d$ from the quantum channel data to a random vector of the same size later used to build the raw secret key.
The recovered information is then larger for a large dimension $d$.
The technique exploits algebraic structures that simplify the mapping for dimensions 1, 2, 4 or 8, while other dimensions require to build a random transformation~\cite{Jouguet2011}.

In this work, we focus primarily on multidimensional reconciliation in high-dimensions, \emph{i.e.} beyond dimension 8. The  aim is to bring together critical elements currently scattered throughout the literature, while providing an educational overview of the fundamentals of reconciliation in CV-QKD. In order to illustrate our developments and actively contribute to the community, we have designed a dedicated open-source software, \HDirac~\cite{hdirac2026}, which implements the most advanced approaches in multidimensional reconciliation, allowing us to simulate a CV-QKD channel with high multidimensional reconciliation. \HDirac closely integrates with \Affect~\cite{Cassagne2019}, a reference open-source library for the fast evaluation of error correction schemes. This allows for different reconciliation methods to be tested in a reproducible, efficient, and transparent framework. The source codes are freely accessible and easily modifiable, so that it is possible to experiment with various error correction codes without having to worry about the implementation details of multidimensional reconciliation. This article also aims to promote closer ties between the CV-QKD and error-correcting code communities by providing detailed educational explanations to facilitate mutual understanding. In addition, the presented simulation results highlight various behaviours that have not been documented before.

Our work complements other recent contributions from the research community. In particular~\cite{Cil2024Qcrypt} is a stand-alone open-source implementation of octonion-based reconciliation; \cite{Gumus2024} targets discrete modulations over a free-space link, but also contains a clear introduction to high-dimensional reconciliation; \cite{Yang2023} is a general review of reconciliation for CV-QKD; and~\cite{Dai2025} proposes an alternative construction of high-dimensional reconciliation that reduces the communication overhead on the classical channel.

First, we present in Section~\ref{sec:protocol} a simplified version of a CV-QKD protocol, and discuss the important concepts required to work on reconciliation. In Section~\ref{sec:virtual}, we describe the construction of the virtual channel of multidimensional reconciliation. In Section~\ref{sec:fec} we discuss various schemes used to combine linear error correction codes with multidimensional construction, thus completing the reconciliation process. In Section~\ref{sec:simulation}, we present and discuss results obtained from open-source LDPC matrices representative of the state of the art~\cite{Cil2024Qcrypt}, in order to compare performance across the different dimensions and configurations studied. We finally conclude this work.

\section{CV-QKD Protocol Overview}
\label{sec:protocol}

We give here a simplified description of a CV-QKD protocol, with the purpose of providing the necessary background to understand the reconciliation process. The description focuses on the prepare-and-measure (PM) protocol with heterodyne detection, reverse reconciliation and Gaussian modulation. Note however that the results of this work are easily adapted to entanglement based protocols, homodyne detection and direct reconciliation. Other modulation formats would usually require different approaches for reconciliation. For completeness, we briefly discuss notable protocol alternatives after the protocol description.

Following the tradition in cryptography, we refer to the two parties involved in the protocol as Alice and Bob, and to the eavesdropper as Eve. Alice and Bob can communicate via a physical quantum channel, which may be manipulated by Eve, but also on a classical errorless authenticated channel: Eve can listen to the classical channel but is prevented from impersonating Alice or Bob.

A QKD protocol must be both secure and robust. \emph{Security} ensures that, regardless of Eve's actions, any key that is established is correct and secret. \emph{Robustness}, also known as completeness, ensures that in the absence of Eve, the protocol generates a key with nonzero probability. This setting is referred to as the \emph{honest} scenario. While the security proof must consider the general case, including arbitrary actions by Eve, theoretical analyses typically establish a key rate that remains positive when Eve's action is sufficiently weak. The practical design and optimization of reconciliation naturally focus on the honest and weak-Eve scenarios.

The individual steps of the protocol have names and general functions common to most prepare-and-measure QKD protocols. They are illustrated in Figure~\ref{fig:protocol_overview} and described below:
\begin{figure}
    \centering
    \includegraphics[width=1\linewidth]{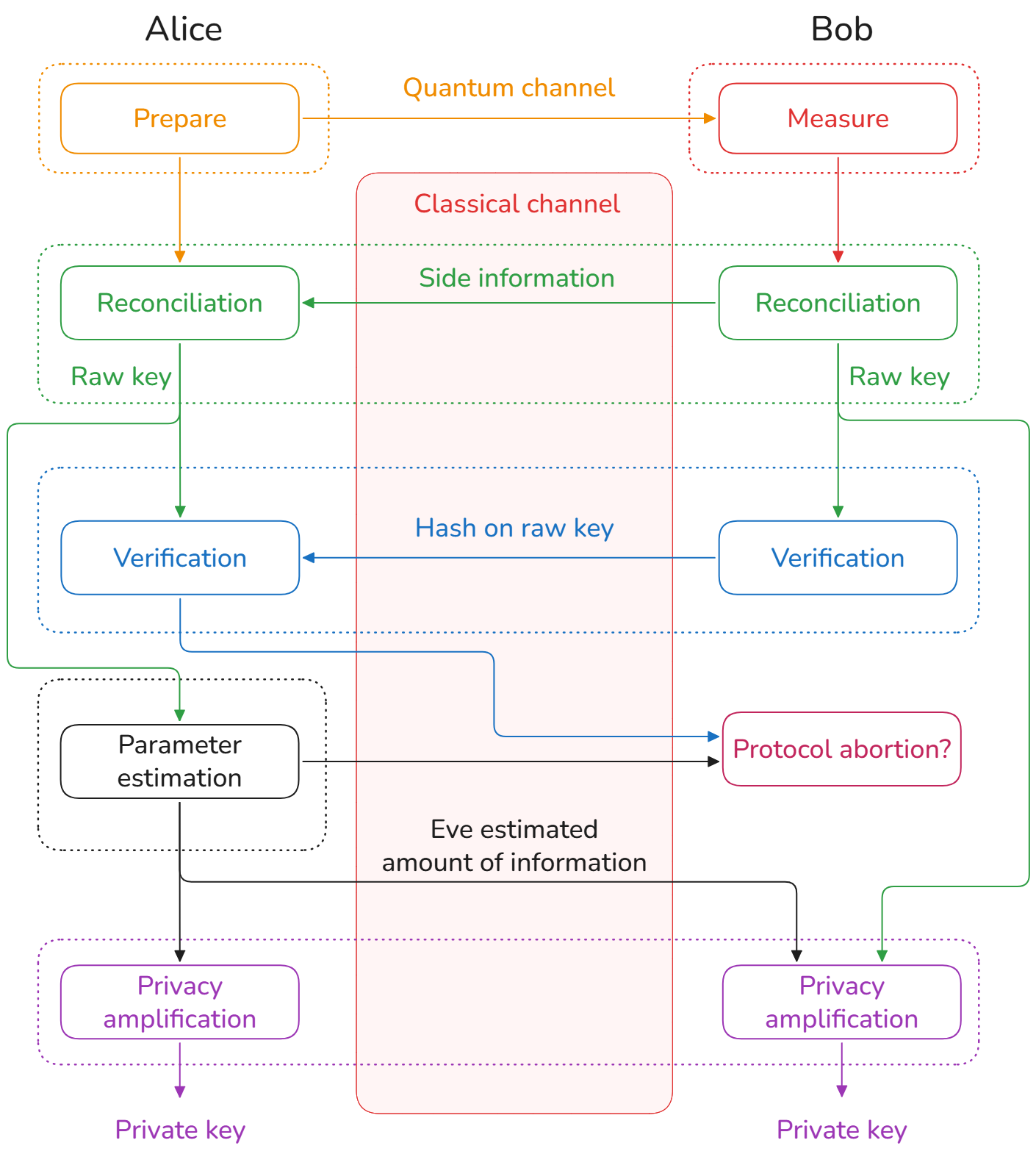}
    \caption{Overview of the general steps of a prepare-and-measure QKD protocol.}
    \label{fig:protocol_overview}
\end{figure}

\begin{enumerate}
    \item \emph{Prepare:} Alice randomly generates, according to a Gaussian distribution \(\mathcal N (0, \sigma_a^2)\)\footnote{Noting \(\mathcal N (0, \sigma^2)\) the normal distribution with mean zero and standard deviation \(\sigma\).}, a QKD \emph{frame} of numbers $a_i$ in $\mathbb{R}$ with $i$ in \(1..L\). She encodes $a_i$ in pairs \((x_k, p_k) = (a_{2k-1}, a_{2k})\) for $k$ in $1..L/2$ into complex numbers \( \alpha_k = (x_k + \mathfrak{j}\cdot p_k)/2 \) where \( \mathfrak{j} \) represents the imaginary unit. Alice modulates a laser pulses in the coherent states \(|\alpha_k \rangle\), then transmits the laser pulses to Bob via the quantum communication channel.

    \item \emph{Measure:} Bob performs a heterodyne\footnote{In this context, \emph{homodyne} usually means detection of a single quadrature, while \emph{heterodyne} usually means simultaneous detection of both quadratures. In laser telecommunications, both are referred to as \emph{coherent detection}.} measurement of the received state: \emph{i.e.} the received light is split in a 50/50 beam splitter, and the two outputs are used to measure, respectively, the two quadratures. The measurement results, arranged in the same order as Alice\footnote{Noting $(\mathtt{x}_{k}, \mathtt{p}_{k})$ the values measured by Bob, $(\mathtt{x}_{k}, \mathtt{p}_{k}) = (b_{2k-1}, b_{2k})$.}, are noted $b_i$.

          Lost light is assumed to be accessible to Eve, and quantum physics dictates that any measurement performed on the transmitted light likely introduces additional noise.

          For the \emph{honest} scenario, we assume the following linear and lossy channel model:

          \begin{equation}
              \label{eq:channel}
              b_i = t \cdot a_i + z_i
          \end{equation}

          where

          \begin{itemize}
              \item \(T = t^2 < 1\) is the transmittance, including Bob's device loss,
              \item \(z_i \sim \mathcal N (0, \sigma_z^2)\) is the total noise (quantum vacuum noise, device imperfections, and channel noise), independent of $a_i$. The channel signal-to-noise ratio is defined as $\text{SNR} = T \sigma_a^2 / \sigma_z^2$.
          \end{itemize}

    \item \emph{Reconciliation:} At this stage, Alice and Bob share random correlated data. Together, they use error correction techniques to build the same common bit string (called the \emph{raw key}). In \emph{reverse} (respectively \emph{direct}) reconciliation, the original bit string must be built purely from Bob's' (resp. Alice's') quantum channel data and a local random number generator; additional data may be revealed by Bob (resp. Alice) on an authenticated channel to enable an efficient error correction by Alice (resp. Bob).

    \item{\emph{Verification:}} Correction is complemented by a verification step to ensure that the bit strings are identical (otherwise, the protocol is aborted). For ITS, this is usually done by disclosing and comparing the output of a function chosen randomly from a universal hash family\footnote{The function is chosen randomly by either Alice or Bob. The choice is then revealed on the classical channel.}.

    \item{\emph{Parameter Estimation:}} A bound is estimated on the amount of information that Eve could maximally hold on the raw key. If Eve could have acquired more information than mutually available to Alice and Bob, the protocol is aborted.

          This estimation is built on a comparison of Alice and Bob data and the fact that any measurement by Eve on the channel would likely introduce a disturbance. In CV-QKD this means that worst-case estimates are computed on the channel parameters: for the most simple protocols, directly the channel transmittance and added noise. Parameter estimation may be performed before reconciliation and may involve the disclosure (and discarding) of a random subset of the quantum channel data.

    \item{\emph{Privacy Amplification:}} This final step involves applying a hash function to the raw key. This process allows Alice and Bob to obtain a perfectly secret bit string from their partially secret raw key. Again, for ITS, universal hashing is generally used.
\end{enumerate}

For the purpose of reconciliation, we can safely assume the model of equation~(\ref{eq:channel})---also called additive white Gaussian noise (AWGN) channel:
The \emph{honest} implementation noise is largely dominated by the Gaussian vacuum fluctuations, and other important technical noise sources (such as thermal noise at the detector) are close to Gaussian. On the other hand, in the presence of Eve, the additional noise tolerated by the protocol - typically quantified as an excess noise variance - is usually too small to have a significant impact on error correction\footnote{For stronger Eve's action, the protocol simply aborts: the protocol stays secure and reconciliation becomes irrelevant.}. However, for the theoretical security analysis and the parameter estimation process, it is in general not safe to make any such assumption on the quantum channel.

The equation describing the reverse reconciliation channel, $a_i = b_i/t - z_i/t$, where $b_i$ is the channel input and $z_i$ is not independent of $b_i$, can be conveniently rewritten as\footnote{Since $a_i$ and $z_i$ are independent Gaussian random variables, they are jointly Gaussian. As $b_i$ and $w_i = a_i - \kappa b_i$ are linear combinations of $a_i$ and $z_i$, they are also jointly Gaussian. Choosing $\kappa = \frac{t\sigma_a^2}{T\sigma_a^2+\sigma_z^2}$ makes their covariance vanish. Thus, $b_i$ and $w_i$ are jointly Gaussian with zero covariance, and hence independent.}
\begin{align}
    a_i & = \mathbb{E}[a_i | b_i] + w_i                               \\
    \label{eq:reverse_channel}
        & = \frac{t \sigma_a^2}{T \sigma_a^2 + \sigma_z^2} b_i + w_i,
\end{align}
where $w_i$ is Gaussian, independent of $b_i$, and has zero mean and variance $\sigma_w^2 = \frac{\sigma_a^2 \sigma_z^2}{T \sigma_a^2 + \sigma_z^2}$. This result reduces the reverse channel to an AWGN channel with the same $\text{SNR} = T \sigma_a^2 / \sigma_z^2$. This property does not generally hold for non-Gaussian modulations.

In order to achieve long distances, finite size theoretical analysis reveals that using long frames is essential. The uncertainty of parameter estimation, in particular, must be limited with a large choice of $L$. $L \approx 10^9$ or larger is usually targeted~\cite{Jouguet2013}. This has important consequences in the design of the reconciliation method and code, as discussed in further detail in this work.

\subsection{Protocol Alternatives}

As mentioned earlier, common protocol alternatives are the following:
\begin{itemize}
    \item Direct reconciliation~\cite{Grosshans2002a}, much more intuitive than reverse reconciliation, limits QKD protocols to 50\% channel transmittance (or $\approx$ 15 km of standard telecom fibre): for smaller $T$, because all lost ($1-T$) light is assumed to be freely measurable by Eve, Eve always has potentially more information on Alice data than Bob. CV-QKD was thus considered unpractical until the realisation that reverse reconciliation has no such limitation~\cite{Grosshans2002b}. Direct reconciliation does not introduce more complexity than presented in this work.

    \item While Gaussian modulation benefits from simpler theoretical proofs and is the most strongly established ~\cite{Leverrier2017, Pirandola2024}, discrete modulations (\emph{i.e.} quadrature amplitude modulation (QAM) with discrete constellations) are also being explored. Small constellations would typically require different reconciliation methods, while large constellations tend to approximate the Gaussian case and can use the same reconciliation methods.

    \item First implementations of CV-QKD used homodyne detection: Bob measures one of the two quadratures chosen randomly, the measurement choice is disclosed, and the other quadrature discarded from Alice data. Homodyne detection is now less common in CV-QKD because it usually required Alice to transmit the local oscillator (the reference laser for the detection) through an optical channel, which introduced security vulnerabilities~\cite{Ferenczi2007}.

    \item Prepare and measure QKD protocols typically have an entanglement based (EB) counterpart: an entangled state is first generated, its components distributed to Alice and Bob, and both parties perform a local measurement. From the theoretical security analysis point of view, the EB description is equivalent and thus widely used. An entanglement based implementation although more challenging has its interests, and some more complex EB protocols cannot be reduced to a PM scenario.

\end{itemize}

\subsection{Secret Key Rate}

The secret key rate (\( \textit{SKR} \)) established between Alice and Bob is easier to express first in the theoretical asymptotic regime, \emph{i.e.} assuming infinitely large QKD frames (as defined in the \emph{Prepare} step of the protocol, here $L \to \infty$):

\begin{equation}
    \textit{SKR}_{\infty} = p_{\textit{qkd}} \cdot (\beta I_{AB} - \chi).
    \label{equa:key_rate_equation}
\end{equation}

\( I_{AB} \), \( \chi \) quantify maximal available information:

\begin{itemize}
    \item \( I_{AB} \) is the classical mutual information between Alice and Bob---for additive white Gaussian noise, the mutual information is maximal (Shannon capacity) for Gaussian modulation and $ I_{AB} = \frac{1}{2} \log_2(1 + \text{SNR}) $.
    \item \( \chi \), the Holevo bound, is the maximum amount of information that Eve could have on the raw key. It is computed from the result of parameter estimation and is more precisely noted \( \chi_{BE} \) for reverse reconciliation (or \( \chi_{AE} \) for direct reconciliation).
          See~\cite{Fossier2009} for asymptotic expressions of $\chi$.
\end{itemize}

The parameters \( \beta \), \( p_{\textit{qkd}} \) quantify imperfections of the reconciliation process:

\begin{itemize}
    \item \( \beta \leq 1 \) is the reconciliation efficiency, defined as the quantity of actually retrieved information divided by the Shannon bound $I_{AB}$ (see also Section~\ref{sec:fec}),
    \item \( p_{\textit{qkd}} \) is the probability, given a specific level of noise, that reconciliation corrects all errors (linked to the Frame\footnote{Note that so far we introduced a single notion of frame, the same for the overall QKD protocol and reconciliation.} Error Rate by \( 1 - \text{FER}_{qkd} \)).
\end{itemize}

At this level, while \( p_{\textit{qkd}} \) is only a scaling factor, $\beta$ is critical and should be close to 100\%: At the system performance limits, \( I_{AB} \) and \( \chi \) are of similar scale such that a small degradation of $\beta$ is enough to shift from a positive to a negative result (without secret key).

Two different unit choices are popular for \( \textit{SKR} \), \( I_{AB} \) and \( \chi \): either bits per laser pulse (with a total of $L/2$ laser pulses), more common and preferred when studying the physics (implementation or theory), or bits per quadrature ($L$ quadratures) preferred when studying the post-processing which directly works on the list of quadrature values. We chose this latter convention in this work. Care should be taken when comparing formulas from different references or for homodyne/heterodyne protocols. Another naming convention is bits per \emph{complex} symbol as opposed to  bits per \emph{real} symbol.

The actual protocol has finite $L$, and the theoretical $\textit{SKR}$ typically requires the following modifications (depending on the variant of protocol and security proof---see \emph{e.g.}~\cite{Leverrier2010, Leverrier2017, Pirandola2024}):
\begin{itemize}
    \item The Holevo bound $\chi({\epsilon_{PE}, L_{PE}})$ now depends on a small security parameter $\epsilon_{PE}$ and the number of symbols $L_{PE}$ used for parameter estimation. As mentioned above, long distances mandate precise estimation and thus large $L_{PE}$.
    \item The subtraction of an information penalty $\Delta(\epsilon, L)$ from $\beta I_{AB} - \chi$ in (\ref{equa:key_rate_equation}). $\epsilon$ is a small and composite security parameter (richer theoretical analysis includes more components to $\epsilon$). $\Delta$ has a dominant contribution in $1/\sqrt{L}$  and a typically negligible contribution in $1/L$ linked to the usage of universal hashing for privacy amplification.
    \item While the \textit{SKR} is useful for theoretical and system analysis, the protocol implementation doesn't use a rate, but rather a number of bits, \emph{i.e.} quantities proportional to $L \cdot \textit{SKR}$.
          $\beta I_{AB}$ is replaced by the entropy of the raw key (either via a worst-case estimation, or the exact value if the raw key is generated by an ideal random number generator), minus the number of bits of this raw key revealed on the classical channel (for reconciliation and verification). This quantity is then reduced further by $L \cdot \chi$ and $L \cdot \Delta$ to give the length of the secret key $l$. Finally, $l$, $L$ and the raw key are used as the inputs of privacy amplification.
\end{itemize}

\subsection{Subframing}

Up to now, we assumed that reconciliation works on the full frame. However, the target \emph{QKD frame} size (\emph{e.g.} $L \approx 10^9$) is larger than practical \emph{error correction frame} size (up to $\approx 10^7$). The main QKD frame can thus be divided into smaller subframes. The subframing strategy, mostly unmentioned in the current literature, has security consequences and must be chosen with care. While the security analysis of subframing in general is ongoing research work~\cite{Johnson2017, Pirandola2024, Jaksch2024}, we describe here three simple methods. The third method is included because it appears to be implicitly assumed in a significant part of the current literature. However, to the best of our knowledge, no proof of its validity has been provided.

We assume the QKD frame divided evenly in $\zeta$ error correction (EC) subframes of length $n$ ($L = \zeta \cdot n$). The code has a probability $p_{\textit{ec}} = 1 - \text{FER}_{\textit{ec}}$ to correct a subframe without error.

\emph{Strict subframing:} In this first method, the whole QKD frame (composed of $\zeta$ subframes) is viewed by the protocol as a single reconciliation frame: verification is performed with a universal hash on the full frame and the protocol aborts if error correction failed on a single subframe.
In this case, equation (\ref{equa:key_rate_equation}) becomes (\ref{eq:skr_strict}) and is valid without further security analysis. Also $p_{qkd} = (p_{ec})^\zeta \approx 1 - \zeta \text{FER}_{ec}$, where the approximation is for small $\text{FER}_{ec}$. The tolerance to a nonzero $\text{FER}_{ec}$ or to small subframes (large $\zeta$) is therefore quite limited: $\text{FER}_{ec}$ should be at most of order $1/\zeta$; otherwise, the probability of protocol success becomes impractically low.

\begin{equation}
    \label{eq:skr_strict}
    \textit{SKR}_{\text{strict}} = (p_{\textit{ec}})^{\zeta} \cdot (\beta I_{AB} - \chi)
\end{equation}

For the two other strategies, additional verification is performed at the subframe level.
For each subframe, Alice estimates whether the decoding was successful and publicly reveals this information. This estimation can be based on the decoder output itself (\emph{e.g.} Did the decoder converge to the correct syndrome?---more in Section~\ref{sec:fec}), and complemented via hashing (Bob discloses a regular hash computed for each subframe. Alice applies the same hash function and compares the result. The number of bits of the revealed hashes is subtracted from the final key length.). For simplicity, we assume here that this estimation never fails: in practice a reasonable hash size enables a very small probability of failure while revealing a small amount of data.

\emph{Reveal-on-error subframing:} For this subframing method, if a subframe was not decoded correctly, Bob reveals the corresponding raw key data. From the protocol point of view, the whole QKD frame is unchanged and verification is performed on this full QKD frame (with universal hashing). Because raw key data are revealed, the key rate is now of the form
\begin{equation}
    \label{eq:skr_reveal}
    \textit{SKR}_{\text{reveal}} = p_{\textit{ec}} \beta I_{AB} - \chi .
\end{equation}
$p_{\textit{ec}}$ now plays a similar role as $\beta$: only a small deviation from 100\% is acceptable. However, the result can be better than strict subframing\footnote{More precisely $\textit{SKR}_{\text{reveal}}>\textit{SKR}_{\text{strict}}$ when $\frac{p_{\textit{ec}} - p_{\textit{ec}}^\zeta}{1 - p_{\textit{ec}}^\zeta} \beta I_{AB} > \chi$. In the limit of large $\zeta$, this reduces to $p_{\textit{ec}} \beta I_{AB} > \chi$ or equivalently $\textit{SKR}_{\text{reveal}}> 0$.}.
Here, $\text{FER}_{\textit{qkd}} = (\text{FER}_{\textit{ec}})^\zeta$: reconciliation fails only when error correction fails on all subframes.

\emph{Discard-on-error subframing:} In this case, the erroneous subframes are discarded from the raw key. The inputs of verification and privacy amplification are thus post-selected depending on Alice error correction results. The hypothesis of reverse reconciliation is broken and a richer security proof is required. One can hope for an \textit{SKR} of the form
\begin{equation}
    \label{eq:skr_discard}
    \textit{SKR}_{\text{discard}} = p_{\textit{ec}} (\beta I_{AB} - \chi - \delta_{\text{discard}}).
\end{equation}
An overall $p_{\textit{ec}}$ factor reflects the fact that a portion of the subframes is discarded; now even a small value is tolerable. This comes at the cost of an additional penalty $\delta_{\text{discard}}$ in the information balance.
The validity of this equation depends on the security framework and is the subject of ongoing research, beyond the scope of this work~\cite{Johnson2017, Pirandola2024, Jaksch2024}.

\subsection{Tuning the Protocol SNR}
Given a CV-QKD implementation on a specific link, the main tunable parameter is Alice modulation variance $\sigma_a^2$ which directly affects the SNR at the receiver. Choosing a suitable $\sigma_a^2$ is a system-level engineering problem and the resulting SNR is central to the reconciliation design. We give here a simplified description: The main inputs are the target security parameter $\epsilon$ ($10^{-10}$ here), the achievable frame size $L$, the excess noise measured on the physical link as a function of SNR, and the achieved $\beta$ as a function of SNR. The SNR is then chosen to maximize the key rate. For concreteness, typical values are given in Table~\ref{tab:numbers}.

Improvements cannot be considered in isolation and must be coordinated across parameters. For example, starting from a balanced system, a one‑order‑of‑magnitude improvement in excess noise yields substantial benefits when combined with a two‑order‑of‑magnitude increase in frame size. As illustrated in the Table~\ref{tab:numbers}, the higher the system performance, the lower the required SNR: hence the extended research by the CV-QKD community on low SNR reconciliation. An improved $\beta$ is also linked to a small increase of optimal SNR.

\begin{table}[htbp]
    \centering
    \caption{Example system parameters, achievable loss and optimal SNR. The excess noise is given as measured by Bob in shot noise units. Achievable loss are often converted to distance in standard telecom single mode fibre with 0.2 dB/km loss.}
    \label{tab:numbers}
    \begin{tabular}{lcccc}
        \midrule
        $L$               & $10^{6}$  & $10^{8}$  & $10^{8}$  & $10^{10}$ \\
        Excess noise      & $10^{-2}$ & $10^{-3}$ & $10^{-3}$ & $10^{-4}$ \\
        $\beta$           & 90\%      & 90\%      & 95\%      & 95\%      \\
        \midrule
        Achievable losses & 3 dB      & 11 dB     & 12 dB     & 16 dB     \\
        Optimal SNR       & 0.8       & 0.08      & 0.10      & 0.035     \\
        \bottomrule
    \end{tabular}
\end{table}

\section{Virtual Channel}
\label{sec:virtual}

In this section, we describe the construction of the virtual channel in dimension $d$, focusing on a subframe of size $n$ and assuming that $n$ is a multiple of $d$. The full picture is then completed in Section~\ref{sec:fec} and Figure~\ref{fig:ldpc_dec_schemes}, adding the usage of error correction codes. We succinctly discuss the usage of Cayley-Dickson algebras for $d \in \{1, 2, 4, 8\}$ and proceed with a more detailed discussion for high dimensions.

The process starts with the generation of the virtual channel input:
Bob generates the raw key as a random binary string, and from the raw key derives the codeword $\vec{c} \in \{0, 1\}^n$. We will see in Section~\ref{sec:fec} how the codeword is derived from the raw key. Bob then builds\footnote{In this article and Figure~\ref{fig:ldpc_dec_schemes} we use the common "BPSK" convention mapping bits $x \in \{0,1\}$ to $(-1)^x \in \{1,-1\}$.} $\vec{u} \in \{1, -1\}^n$ via $u_i = (-1)^{c_i}$.

The output is then built with the quantum channel data arranged in blocks of size $d$. A vector $\vec{x} \in \Omega^n$ (where $\Omega$ is either $\mathbb{R}$ or $\{1, -1\}$) is partitioned with the following notation, for $i$ in $(1, \ldots, n/d)$ :

\[ {}_d\vec{x}_i = (x_{(i-1)d+1}, \ldots, x_{id}) \in \Omega^d \]

The prescript $d$ denotes the fact that the corresponding vector lives in dimension $d$.

\subsection{Cayley-Dickson Algebras}

Up to $d=8$, Cayley-Dickson algebras allow for a simplified construction of the virtual channel. These four algebras are the real and complex numbers ($d=1$ and $2$), the quaternions ($d=4$) and the octonions ($d=8$) which in particular have a multiplicative inverse: for a non-zero vector ${}_d\vec{x} \in \mathbb{R}^d$ there is a unique inverse ${}_d\vec{x}^{-1}$ satisfying
\[
    {}_d\vec{x} \cdot {{}_d\vec{x}_{}}^{-1} = {{}_d\vec{x}_{}}^{-1} \cdot {}_d\vec{x} = {}_d1,
\]
where $\cdot$ is the multiplication in this algebra and ${}_d1$ is the multiplicative identity.

The output of the virtual channel is obtained with these steps, with $\vec{a}$ and $\vec{b}$ defined in equation (\ref{eq:channel}):
\begin{enumerate}
    \item Bob computes a vector $\vec{r} \in \mathbb{R}^n$, with for each $i \in (1, \ldots, n/d)$: $${}_d\vec{r}_i = {}_d\vec{u}_i \cdot {}_d\vec{b}_i.$$
    \item Bob discloses $\vec{r}$ on the classical channel.
    \item Alice computes the output $\vec{v}$ of the virtual channel: $${}_d\vec{v}_i = {}_d\vec{r}_i \cdot {{}_d\vec{a}_i}^{-1}.$$
    \item $\vec{v}$ is a noisy and rescaled version of the raw key as we show below. Alice can thus proceed with error correction (Section~\ref{sec:fec}) to retrieve the raw key.
\end{enumerate}

$\vec{v}$ can be rearranged as:

\begin{align}
    {}_d\vec{v}_i
     & = ({}_d\vec{u}_i \cdot {}_d\vec{b}_i) \cdot {{}_d\vec{a}_i}^{-1}                           \\
     & = ({}_d\vec{u}_i \cdot (t \cdot {}_d\vec{a}_i + {}_d\vec{z}_i)) \cdot {{}_d\vec{a}_i}^{-1} \\
     & = t \cdot {}_d\vec{u}_i + ({}_d\vec{u}_i \cdot {}_d\vec{z}_i) \cdot {{}_d\vec{a}_i}^{-1}
\end{align}

Note that the algebra is distributive but \emph{not} associative. It does, however, satisfy weaker forms of associativity. In the last line, we used distributivity and $(xy)y^{-1} = x(yy^{-1}) = x$.

\subsection{High-Dimensional Reconciliation}
\label{high_dimensional_rec}

In higher dimensions, the virtual channel output is computed as follows:

\begin{enumerate}
    \item Bob generates random orthogonal transformations ${}_{d}\R_i$ of $\mathbb{R}^d$ with the constraint that ${}_d\vec{b}_i$ is mapped to ${}_d\vec{u}_i$ up to a scaling factor:  ${}_{d}\R_i({}_d\vec{b}_i) = \|{}_d\vec{b}_i\| \frac{{}_d\vec{u}_i}{\sqrt{d}}$.
    \item Bob discloses $\{{}_{d}\R_i\}$ and $\{\|{}_d\vec{b}_i\|\}$ on the classical channel.
    \item Alice computes the output $\vec{v}$ of the virtual channel: \begin{equation}
              \label{eq:virtual_output}
              {}_d\vec{v}_i = {}_{d}\R_i({}_d\vec{a}_i).
          \end{equation}
    \item $\vec{v}$ is again a noisy and rescaled version of the raw key.
\end{enumerate}

This process involves finding the orthogonal transformations that map the vectors ${}_d\vec{b}_i$ to ${}_d\vec{u}_i$, but picking deterministically any transformation would reveal too much information to Eve. The transformation must thus be picked randomly. \cite{Jouguet2011} propose two constructions that we present in Appendix~\ref{appendix:householder}. The first is in complexity $O(d^3)$ and can use the $QR$ decomposition readily implemented in popular linear algebra libraries. The second is a modified version of Householder method for $QR$ decomposition, better suited to our particular problem: it has an improved $O(d^2)$ complexity.

More explicitly, we show in Appendix~\ref{appendix:virtual_theory} how the virtual channel output distribution given ${}_d\vec{u}_i$ and the public ${}_{d}\R_i$ and $\|{}_d\vec{b}_i\|$ can be written as $d$ BIAWGN channels:

\begin{align}
    {}_d\vec{v}_i \mid {}_d\vec{u}_i \sim \mathcal{N}\left(\frac{\text{SNR}}{1+\text{SNR}} \frac{\|{}_d\vec{b}_i\|}{t} \frac{{}_d\vec{u}_i}{\sqrt{d}}, \frac{\text{SNR}}{1+\text{SNR}} \frac{\sigma_z^2}{T} \mathbb{I}\right).
\end{align}

The input of a modern coding decoder is generally a binary log-likelihood ratio (LLR): For a channel input $x \in \{0, 1\}$ and measured value $y$,
\begin{equation}
    \label{eq:llr_definition}
    \text{LLR} = \text{ln} \frac{P(y | x = 0)}{P(y | x = 1)}.
\end{equation}
In particular, for a BIAWGN channel with  $y = (-1)^x + \mathcal{N}(0, \sigma^2)$, $\text{LLR}_{\text{BIAWGN}} = 2 y / \sigma^2$.

We can therefore write after proper scaling:

$${}_d\overrightarrow{\text{LLR}}_i = \frac{2 t \|{}_d\vec{b}_i\|}{\sqrt{d}} \frac{{}_d\vec{v}_i}{\sigma_z^2}.$$

Alice can use the full vector $\overrightarrow{\text{LLR}}$ as the input of her decoder.

We also show in the Appendix~\ref{appendix:virtual_theory} that the mutual information of the virtual channel is
$$
    I_d = \mathbb{E}_{\lambda \sim \chi_{d}^2}\left(I_{\text{BIAWGN}}\left(\frac{\lambda}{d} \text{SNR}\right)\right)
$$
where $I_{\text{BIAWGN}}(1/\sigma^2)$, is the mutual information of the BIAWGN channel for a given signal-to-noise ratio $1/\sigma^2$.

\begin{figure}[ht]
    \centering
    \includegraphics[width=1\linewidth]{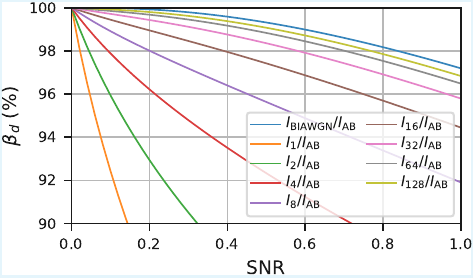}
    \caption{Upper bound on the reconciliation efficiency of the virtual channel as a function of SNR.}
    \label{fig:virtual_theory}
\end{figure}

Figure~\ref{fig:virtual_theory} displays the numerically computed $\beta_d = I_d / I_{AB}$ as a function of SNR, showing how the virtual channel mapping intrinsically loses information from the physical channel, but with negligible penalty for large enough $d$ or small enough SNR.

Independently of our specific virtual channel mapping, there is also an information loss from the approximation of the physical channel---which may carry an infinite amount of information per symbol at large SNR---by a binary input channel---which naturally saturates at 1 bit per symbol. This is illustrated by the plot of $I_{\text{BIAWGN}}/I_{AB}$ in Figure~\ref{fig:virtual_theory}, materializing the limit of multi-dimensional reconciliation for large $d$.

\begin{figure*}[!ht]
    \centering
    \includegraphics[width=1\linewidth]{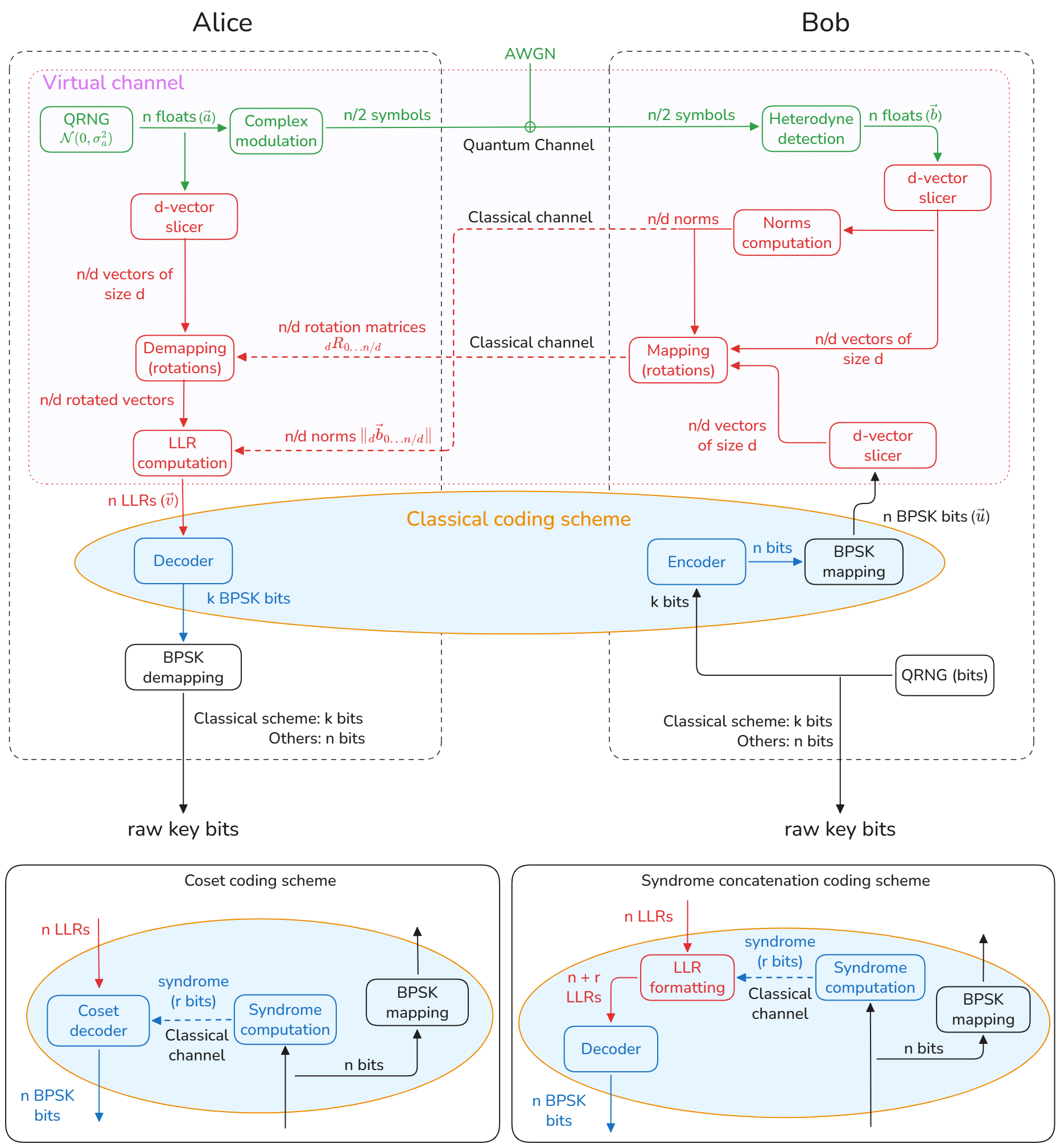}
    \caption{Linear block coding schemes for multidimensional reverse reconciliation. The physical layer (prepare and measure steps of the protocol) is highlighted in green, the multidimensional mapping in red, (virtual) channel coding in blue, and bit generation/mapping in black. QRNG: Quantum Random Number Generator (or any true random number generator). Refer to sections \ref{sec:virtual} and \ref{sec:fec} for a description of the individual components.}
    \label{fig:ldpc_dec_schemes}
\end{figure*}

\section{Error Correction}
\label{sec:fec}

Once the virtual channel data are built, Alice and Bob can adapt and use standard error correction techniques. Several types of error-correcting codes have been explored for CV-QKD: polar codes~\cite{Jouguet2014, Cao2023}, raptor codes~\cite{Shirvanimoghaddam2015, Zhou2019}, Multi-Edge Type LDPC codes~\cite{Jouguet2011, Hossein2021}, Hadamard LDPC codes~\cite{Xing2025}, spatially coupled LDPC codes~\cite{Jiang2018}, protograph based LDPC codes~\cite{Gumus2021, Cil2024} repeat-accumulate LDPC codes~\cite{Johnson2015}, also, but to a lesser extent spinal codes~\cite{Wen2020}. All these codes can have different variants in terms of construction, decoding algorithm, and hardware or software implementation (for example LDPC codes can be quasi-cyclic~\cite{Milicevic2018}, rate adaptative~\cite{Liu2014}).

In this section, we describe various schemes---illustrated in Figure~\ref{fig:ldpc_dec_schemes}---allowing to use error correction codes on the virtual channel, restricting for simplicity the discussion to binary linear block codes. This includes LDPC codes which are the most studied for CV-QKD. See for example~\cite{Lodewyck2006} for early insights on the schemes described here. Later, in the simulation section, we use LDPC codes provided by~\cite{Cil2024Qcrypt}, and a popular decoder when both complexity and reconciliation efficiency matter: namely Belief Propagation (BP) with the Sum-Product Algorithm (SPA).

\subsection*{Classical Coding}
\emph{Classical coding} relies directly on the standard usage of linear codes. Consider a message \( \vec{m} \in \{0, 1\}^k \); channel coding adds redundancy to this message before transmission on a noisy channel. Linear block codes define a binary \emph{generator} matrix \( G \) of size \( k \times n \). Encoding consists in multiplying the (transposed) generator matrix by the message to obtain the codeword \(\vec{c} =  G^T \vec{m} \in \{0, 1\}^n \). The \emph{code rate} is defined as the density of information $R=k/n$ in a codeword.

Decoding, performed on receiver side, aims to detect and correct bit errors introduced by disturbances during the transmission of the codeword. A matrix called the \emph{parity check matrix} \( H \) of size \( (n - k) \times n \), is generally used. It represents a set of parity equations \( H \vec{c} = \vec{0} \) imposed during the encoding and is derived from $G$ by solving \( H G^T = 0 \). A noisy codeword will usually have a non-zero \emph{syndrome} $\vec{s} = H \vec{c} $

The decoder then generally works on LLRs (as defined in equation (\ref{eq:llr_definition})). Using this "soft" information, the decoder attempts to satisfy the parity equations. This is typically done via an iterative algorithm, called \emph{message passing}, or \emph{belief propagation} which gradually updates the LLRs according to the code constraints. At each iteration, the confidence levels evolve to make the received word increasingly consistent with the parity equations.
This process can be interpreted as an attempt to make the estimated codeword converge towards a zero syndrome, \emph{i.e.} towards a situation where \( H\vec{c} = \vec{0} \).
Once the algorithm has converged, the decoder makes a hard decision on the updated LLRs to produce a final binary estimate of the codeword. The correct message must then be computed from the codeword.

In our context, a code is chosen in advance. Bob generates randomly the raw key in \( \{0, 1\}^k \) which is simply used as the message $\vec{m}$ (see Figure~\ref{fig:ldpc_dec_schemes}).

The overall reconciliation efficiency is given by: $$\beta I_{AB} = \frac{\text{Raw key bits}}{\text{Virtual channel bits}} = \frac{k}{n}.$$ Hence, $$\beta = \frac{R}{I_{AB}}.$$

Note that this formula is often presented as a definition of $\beta$. Here, we defined $\beta$ as an information ratio independent of the coding technique. We show in this section how $\beta$ is related to the code rate with the same formula $\beta = R/I_{AB}$ for all the presented schemes.

\subsection*{Coset Coding}
In \emph{coset coding}, Bob directly generates the raw key as a random bit string $\vec{c}$ of size \( n \) and computes the non-zero syndrome \( \vec{s} = H \vec{c} \). $\vec{c}$ is not indeed a codeword of our linear code with parity check matrix $H$, but a codeword from the \emph{coset} defined, for a fixed $\vec{s}$, by $\{ \vec{c} \in \{0, 1\}^n: \vec{s} = H \vec{c} \}$. Bob sends \( \vec{c} \) on the virtual channel and the syndrome \( \vec{s} \) on the classical channel: $r = n - k$ bits of information from the raw key are revealed.

When Alice attempts to decode \( \vec{c} \), she seeks convergence to the parity equations of the coset code, namely \( H \vec{c} = \vec{s} \). This requires a minor, but non-standard, modification of the decoding algorithm.

This method has the key advantage that the encoding process is completely eliminated. In particular, while $H$ is typically a sparse matrix (hence the name Low Density Parity Check code---LDPC), $G$ is a large and dense matrix which is unpractical to build, store and use.

Here, the overall reconciliation efficiency is given by: $$\beta I_{AB} = \frac{\text{Raw key bits - Revealed bits}}{\text{Virtual channel bits}} = \frac{n - r}{n} = \frac{k}{n}.$$ The result is the same as for classical coding: $$\beta = \frac{R}{I_{AB}}.$$

\subsection*{Syndrome-Concatenation Coding}
\emph{Syndrome concatenation} is an alternative---mathematically equivalent---presentation of coset coding, but it will be more familiar to the reader experienced in error correction and in particular systematic codes. It also highlights how a classical decoder can be tricked into performing coset decoding.

From a base code with parity check matrix $H$ of size $r \times n$, a larger systematic code is built by concatenating to a message $\vec{m} \in \{0, 1\}^n$ the "parity bits" $\vec{p} = H\vec{m}$ into a codeword $\vec{c} = \left[\, \vec{m} \mid \vec{p} \,\right] \in \{0, 1\}^{n+r}$. This larger code is in \emph{systematic} form, meaning that the message can be read directly from the codeword (here by selecting the first $n$ bits), and has generator and parity check matrices

\[ G' = \left[\, \mathbb{I}_n \mid H^T \,\right] \text{ and } H' = \left[\, H \mid \mathbb{I}_r \,\right]. \]

For reconciliation, the raw key is $\vec{m}$ and is transmitted on the virtual channel, while the parity bits $\vec{p}$ are transmitted on the errorless classical channel. The full codeword is thus transmitted on a composite channel.

Alice may then use a classical decoder for the code $H'$: the first $n$ LLRs are computed from the virtual channel formula, and the last $r$ LLRs, from the errorless channel, have values $\pm{\infty}$ corresponding to perfect knowledge.

While this technique allows to use a standard decoder without modification, it requires the processing of an $n + r$ LLRs instead of $n$. This is a significant penalty in the context of CV-QKD and its low code-rate regime.

The reconciliation efficiency is now given by:

$$\beta I_{AB} = \frac{\text{Raw key bits - Revealed bits}}{\text{Virtual channel bits}} = \frac{n - r}{n} = \frac{k}{n}.$$ The result is again: $$\beta = \frac{R}{I_{AB}}.$$

Note that it is the rate $R$ of the base code that matters, and not the rate $R'=\frac{n}{n+r}$ of the larger systematic code.

\section{Simulation}
\label{sec:simulation}

\subsection{Simulation Framework}

\subsubsection{\Affect Toolbox}
The open-source simulation software we have developed, \HDirac, is based on \Affect~\cite{Cassagne2019}, which is open source and licensed under the MIT licence. It is dedicated to simulating error correction systems. Written in C++11, it supports a wide range of correction codes. Its architecture is organised around two main abstractions:

\begin{itemize}
    \item \emph{Modules}, which represent computational entities (\emph{e.g.} a modulator or a channel).
    \item \emph{Tasks}, which correspond to operations performed within a module (\emph{e.g.} modulation, demodulation, or addition of Gaussian noise).
\end{itemize}

This architecture is based on \StreamPU~\cite{Cassagne2023}, a domain-specific embedded language dedicated to software-defined radio. It enables the concise description of parallel dataflow graphs and efficiently exploits multicore architectures through pipelining and replication parallelisms.

Our initial implementations relied directly on the built-in decoder using syndrome-concatenation coding. We subsequently developed a new decoder supporting coset decoding (which is not natively available in \Affect), together with less aggressive mathematical approximations---the original decoder is optimised for high-speed operation in standard telecommunications scenarios, albeit at the cost of reduced reconciliation efficiency in the CV-QKD regime. The decoder implements a flooding sum-product algorithm (SPA). The results presented here are obtained using this new decoder for both coset coding and syndrome-concatenation coding.

\subsubsection{\Eigen Library}
\HDirac also relies on the \Eigen library~\cite{Guennebaud2010}, a C++ library for linear algebra that provides decompositions, numerical solvers and algorithms for matrix and vector operations. In our implementation, \Eigen is used to handle linear algebra required in the Householder $O(d^3)$ implementation of the virtual channel.

\subsubsection{LDPC Codes}
We performed simulations using the parity check matrices (\( H \)) published in~\cite{Cil2024Qcrypt}. These are LDPC matrices constructed from protographs, with adaptable code rates, presenting a state-of-the-art \( \beta \)/FER trade-off. Decoding was limited to syndrome concatenation and coset approaches (as generator matrices G were unavailable). These simulations allowed us to compare frame error rates as a function of \( \beta \), for different reconciliation dimensions and for several relevant code rates for CV-QKD with multidimensional reconciliation.

\subsubsection{Simulation Parameters}

Comparing the performance of various codes applied to CV-QKD reconciliation requires specifying a number of parameters. We list here the most important parameters and highlight those which are accessible in \HDirac.

\begin{itemize}
    \item \emph{Type of error correction code} (\emph{e.g.}, LDPC, polar, raptor, ...). \emph{Decoding algorithm} (\emph{e.g.}, horizontal layered SPA, successive cancellation) and its parameters (number of iterations, implementation variants and hardware). \emph{Construction} of the code.

          This work focuses on a specific LDPC code and flooding SPA, but \HDirac can directly be used with other LDPC codes and can easily be adapted to other type of codes thanks to its integration with \Affect.

    \item \emph{Simulated channel} (\emph{e.g.} BIAWGN, virtual for a given $d$, slice reconciliation channel). \HDirac implements the BIAWGN and virtual channels with various constructions and arbitrary dimension $d$.

    \item \emph{Coding scheme} (Accessible in \HDirac: classical, coset, syndrome-concatenation).

    \item \emph{Code rate} $R$ (accessible in \HDirac).

    \item \emph{Codeword size} $n$ (accessible in \HDirac).

    \item \emph{SNR}, or equivalently, reconciliation efficiency $\beta$ (accessible in \HDirac)\footnote{Note that $\beta$ is computed using the realistic CV-QKD Gaussian channel capacity for virtual channels and the BIAWGN channel capacity for BIAWGN simulations.}.
\end{itemize}

Some output of the simulation are input parameters of the protocol:
\begin{itemize}
    \item \emph{Frame error rate} (FER), or more precisely number of frames simulated and number of errors (main result of our simulation).

    \item \emph{Throughput} and \emph{latency} (results of the simulation, with various possible units).
\end{itemize}

As an illustrative example, consider a code with rate \( R \approx 0.1 \) using a coset decoder with a maximum of 500 iterations. For parameters \( d = 64, k = 20000, n = 204800 \text{ and } \beta = 93\% \), the simulation takes approximately one minute (on a single thread) to decode 100 frames, with no observed frame errors.

In the results presented in this paper, we fix the maximum number of SPA iterations to 500, and stop the simulation at the earliest of 100 frame errors or 1000 simulated frames\footnote{This explains why the error area is larger in regions with a high FER.}.

Typical simulator performances are given in Table~\ref{tab:throughput}. The throughput\footnote{The throughput is defined here as the number of information ($k$) bits processed---possibly leaving uncorrected errors---per unit time.} for a virtual channel of dimension \(d = 64\) and coset coding are measured on eight threads (\emph{AMD Ryzen AI 9 HX 370} CPU) with 32 GiB of RAM (quad-channel LPDDR5x). The simulations were performed on \Dalek~\cite{Cassagne2025}, a cluster dedicated to emerging architectures and energy-aware. The results are reported for different values of \( n \), while keeping \( k = 20000 \) fixed ($R \approx$ 0.2, 0.1 and 0.02). Two regimes of \( \beta \) are considered: one corresponding to a frame error rate (FER) of approximately \( 100\% \) (\( \beta = 100\% \)), and another corresponding to a near-zero FER regime (\( \beta = 90\% \)).

\begin{table}[ht]
    \centering
    \begin{tabular}{cc|cc}
        \hline
        \( R \)        & \( n \) & \( \beta = 100\% \) & \( \beta = 90\% \) \\
        \hline

        $\approx 0.2$  & 102400  & 126                 & 281                \\
        $\approx 0.1$  & 204800  & 43                  & 130                \\
        $\approx 0.02$ & 1024000 & 6                   & 25                 \\
        \hline
    \end{tabular}
    \caption{Throughput (kb/s of processed information bits) with a decoder parallelised on eight threads (\( k = 20000 \)).}
    \label{tab:throughput}
\end{table}

\subsection{Simulation Results}

\begin{figure*}[!ht]
    \centering
    \includegraphics[width=1\linewidth]{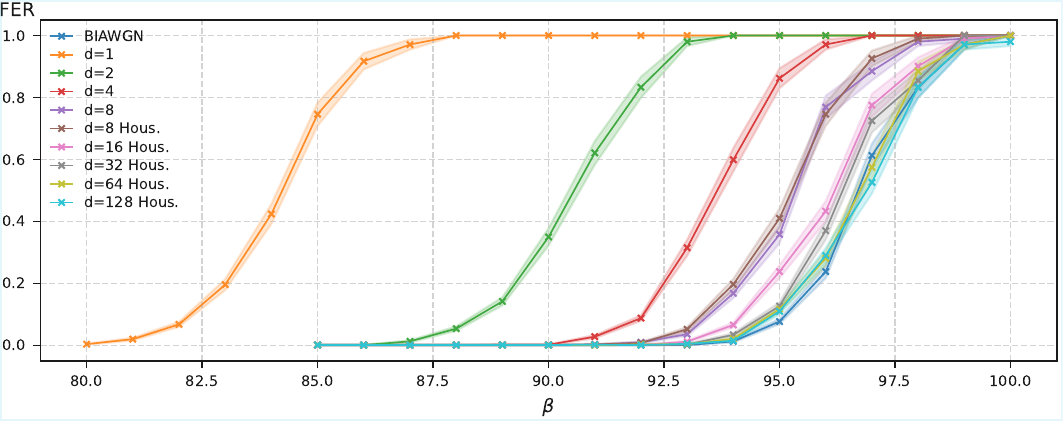}
    \caption{FER as a function of $\beta$ for $R \approx 0.1$ (\( k=20000, n = 204800\)) and coset coding.}
    \label{fig:all_dimensions}
\end{figure*}

Figure~\ref{fig:all_dimensions} shows the coupled behaviour of the two reconciliation parameters of the CV-QKD \( \textit{SKR} \) (Equation (\ref{equa:key_rate_equation})): the reconciliation efficiency $\beta$ and the FER. The figure illustrates the typical transition from the high SNR/low $\beta$ regime with no decoding errors, to the low SNR regime with high error rates. With equations (\ref{eq:skr_strict}) and (\ref{eq:skr_reveal}) the optimal regime corresponds to the highest $\beta$ while maintaining the constraint $\text{FER} \approx 0\%$. Equation (\ref{eq:skr_discard}) allows for higher FER and thus higher $\beta$.

The measurements are performed for various dimensions $d$, showing how increasing the dimension of the virtual channel improves the performance. As we can see, for this code rate $R \approx 0.1$, using $d=8$ (octonions) is already a large ($> 10 \%$) improvement from $d=1$. $\beta$ can be further improved ($1.3 \%$) with $d$ increased to 64, with a result almost indistinguishable from the code performance on the BIAWGN channel.

For $d \leq 8$, Cayley-Dickson algebra are used while for $d \geq 8$, the Householder method is used. No significant difference is observed at $d=8$ for the octonion implementation and the Householder implementation (labelled \emph{Hous.}).

\begin{figure*}[!ht]
    \centering
    \includegraphics[width=1\linewidth]{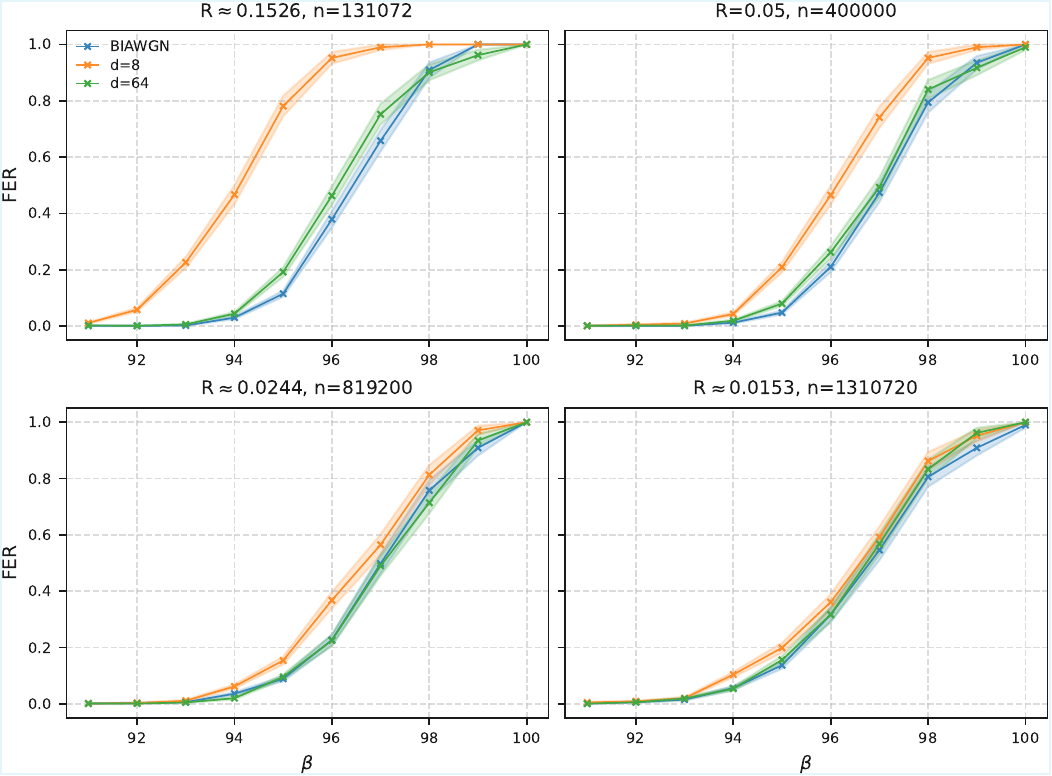}
    \caption{FER as a function of $\beta$ for various $R$, fixed $k=20000$ and coset coding.}
    \label{fig:diff_rates}
\end{figure*}

Figure~\ref{fig:diff_rates} shows similar curves for various code rates $R$ and fixed $k = 20000$. The figure illustrates that as $R$, and thus the SNR, is reduced, the virtual channel tends to a better approximation of the BIAWGN channel, even with low dimension $d$. We can also see that for this set of codes with fixed $k$, the BIAWGN channel curve only marginally depends on $R$.

\begin{figure}[!ht]
    \centering
    \includegraphics[width=1\linewidth]{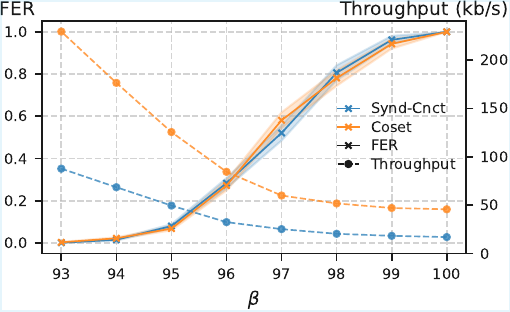}
    \caption{FER and processed information throughput on the BIAWGN channel for $R \approx 0.1$ (\( k=20000, n = 204800\)) and either syndrome-concatenation or coset coding. The decoder was parallelised on eight CPU threads on the \emph{Dalek} cluster, under the same conditions as for Table~\ref{tab:throughput}.}
    \label{fig:comp_synd_cnct_coset}
\end{figure}

Figure~\ref{fig:comp_synd_cnct_coset} compares coset coding and syndrome concatenation coding. It confirms that the error rate is similar for these two methods. However, the information throughput is reduced for syndrome-concatenation coding because a larger amount of LLR nodes must be processed (larger by a ratio $\frac{n+r}{n}$, so $\approx  2$ for our low code rates).

\begin{figure}[!ht]
    \centering
    \includegraphics[width=1\linewidth]{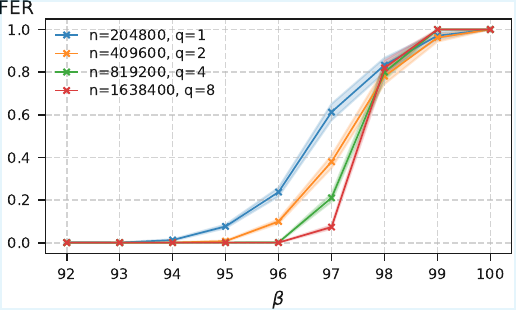}
    \caption{FER on the BIAWGN channel as a function of $\beta$ for $R \approx 0.1$, coset coding and quasi-cyclic expansion of the code size by a factor $q$.}
    \label{fig:qc_comp}
\end{figure}

Finally, Figure~\ref{fig:qc_comp} shows that the number of information bits (\( k \)) has a significant impact on the performance of a LDPC code. Here, the codes from~\cite{Cil2024Qcrypt} were expanded by a factor $q$ with a \emph{quasi-cyclic} construction. As \( k \) increases towards infinity, the "waterfall" curve observed in a \( \beta \) as a function of FER plot becomes steeper. In fact, it pivots around a FER of approximately 80\% here and also for the other code constructions that we have tested. In these examples, increasing the code size allows to achieve a better reconciliation efficiency as soon as the target FER is smaller than 80\%, but at the cost of increasingly unpractical implementations.

\section{Conclusion}
In this work, we have presented a detailed review of multidimensional reconciliation in continuous-variable quantum key distribution. We established that the reverse-reconciliation channel has the same form as its direct-reconciliation counterpart, thereby providing a unified description of both reconciliation directions.
We then focused on high-dimensional constructions relevant to long-distance and low-SNR regimes. We discussed the generalisation from algebraic constructions to arbitrary dimensions, and how it relies on random orthogonal transformations to preserve secrecy while enabling efficient decoding. Building on this virtual channel, we reviewed several practical schemes that allow linear error-correcting codes to be deployed in CV-QKD systems, highlighting their equivalence in terms of reconciliation efficiency and their different implementation trade-offs.
A key contribution of this work is the introduction of \HDirac, an open-source simulation platform designed to be both flexible and reproducible. By integrating tightly with the \Affect library and leveraging recent protograph-based LDPC codes, the simulator enables performance comparisons across dimensions, code rates, and reconciliation schemes. The results confirm that high-dimensional reconciliation can improve performance in the low-SNR regime, while also revealing non-trivial dependencies between the reconciliation dimension, frame error rate, and achievable efficiency. These observations underline the importance of jointly optimizing the reconciliation dimension, coding strategy, and system parameters rather than considering them in isolation.
Beyond simulation, this article aims to lower the barrier between the CV-QKD and error-correction communities by providing explicit derivations, clear terminology, and practical implementation insights. As CV-QKD systems move towards higher symbol rates, longer distances, and real-time operation, reconciliation will remain a central performance bottleneck. Ongoing work in the research community naturally extends towards tighter finite-size security analyses incorporating subframing strategies, improved modelling of the virtual channel noise, hardware-aware decoder optimisations, and the exploration of alternative coding paradigms.
By consolidating existing knowledge and proposing practical tools, we hope that this work will contribute to the continued maturation and deployment of high-performance CV-QKD systems.

\section*{Acknowledgements}

This work was supported by the European Commission under the project QKISS, grant agreement no. 101091448, and the project QSNP, grant agreement no. 101114043. The authors also acknowledge the support of the CIFRE PhD grant of Lucien Martial. The authors would like to thank Philippe Grangier for initiating this collaboration and for fruitful discussions.

\appendix
\section{Householder Methods}
\label{appendix:householder}

We describe here the methods proposed in~\cite{Jouguet2011} to adapt the Householder $QR$ decomposition method, and the statistical analysis of~\cite{Stewart1980} for multidimensional transformations in $O(d^3)$ and $O(d^2)$ complexity.

\subsection{Householder Transformation}

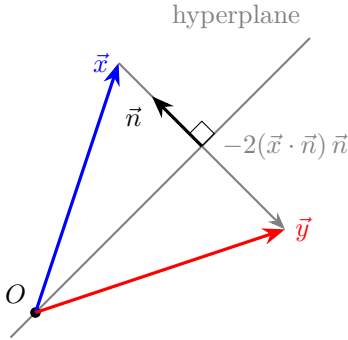
\begin{figure}[ht]
    \centering
    \begin{tikzpicture}[scale=1.1, >=Stealth]
        \coordinate (O) at (0,0);
        \coordinate (X) at (1,3);
        \coordinate (Y) at (3,1);

        \coordinate (M) at (2,2);

        \draw[gray, thick] (-0.3,-0.3) -- (3.3,3.3)
        node[above left, gray] {hyperplane};

        \fill (O) circle (1.8pt) node[above left] {$O$};
        \draw[->, blue, very thick] (O) -- (X) node[left] {$\vec{x}$};
        \draw[->, red,  very thick] (O) -- (Y) node[right] {$\vec{y}$};

        \draw[->, gray, thick] (X) -- (Y)
        node[midway, right=4pt, gray] {$-2(\vec{x}\cdot\vec{n})\,\vec{n}$};

        \draw[->, black, very thick] (M) -- ++(-0.6,0.6)
        node[below left] {$\vec{n}$};

        \draw (M) -- ++(0.15, 0.15) -- ++(-0.15,0.15) -- ++(-0.15,-0.15);

    \end{tikzpicture}
    \caption{Householder Transformation.}
    \label{fig:householder}
\end{figure}

The reflection in $\mathbb{R}^d$ with respect to a hyperplane containing the origin is defined entirely by a unit normal vector $\pm\vec{n}$.
Such a transformation is also called a Householder transformation as is illustrated in Fig.~\ref{fig:householder}.
The transformation from $\vec{x}$ to $\vec{y}$ is given by the formula:

$$\vec{y} = \vec{x} - 2 (\vec{x} \cdot \vec{n}) \vec{n}$$

Note in particular that the complexity of this transformation is $O(d)$, while a general orthogonal transformation has complexity $O(d^2)$.

Reversely, given two distinct vectors $\vec{x}$ and $\vec{y}$ of the same norm, there exists a unique Householder transformation that maps $\vec{x}$ to $\vec{y}$, described by the normal vector $\pm\vec{n} = \frac{\vec{x}-\vec{y}}{\|\vec{x}-\vec{y}\|}$.

If $d=1$, the unique Householder transformation transforms $x$ to $-x$.

For the purpose of this appendix we use a modified definition of a Householder transformation, allowing $\vec{n} = \vec{0}$ to describe the identity.

\subsection{QR Decomposition Method}
\label{subsec:QR}
The first method proposed in~\cite{Jouguet2011} and described in more detail in~\cite{Gumus2024} is the following.

Given two normalised vectors $\vec{b}$ and $\vec{u}$ in $\mathbb{R}^d$, our goal is to build a random orthogonal transformation $\R$ such that $\R(\vec{b}) = \vec{u}$.

\begin{itemize}
    \item Generate a $d \times d$ matrix by drawing randomly each of its elements from \(\mathcal N (0, 1)\). Compute the $QR$ decomposition with a positive diagonal for R (complexity $O(d^3)$) of this matrix. Keep the $Q$ component: $Q$ is a Haar-random orthogonal matrix~\cite{Stewart1980}.
    \item Find $H$, the Householder transformation such that $H(Q.\vec{b}) = \vec{u}$.
    \item Our result is the transformation $\R = H \circ Q$ (where we use $\circ$ to denote composition.).
\end{itemize}

\subsection{Stewart Method}

\newcommand{\lsub}[2]{\prescript{}{#1}{\vec{#2}}}
\newcommand{\lsu}[2]{\prescript{}{#1}{#2}}

In~\cite{Stewart1980}, Stewart describes another method to generate a random orthogonal transformation in dimension $d$. A slightly modified version is the following: For each dimension $i$ from $d$ to $2$, generate a random Householder transformation by drawing a random normal vector $\lsub{i}{n}$ (draw $i$ coefficients from a standard normal distribution, and normalise the resulting vector), and extend this transformation to the whole space. For $i=1$, pick randomly a transformation that multiplies the input by $\pm1$ ($\lsub{0}{n} \in \{0, 1\}$). Compose these $d$ transformations for $i$ from $1$ to $d$.

Noting that this construction is closely related the Householder $QR$ decomposition of a matrix (which uses iteratively Householder transformations for dimensions $d$ to $2$),
Stewart shows that the resulting transformation is uniformly distributed for the Haar measure in the orthogonal group $\mathcal{O}(d)$.
Furthermore, the complexity of this method is $O(d^2)$ (both in time and space).

\subsection{Adding the Constraints of Multidimensional Reconciliation}

The second method of Jouguet \emph{et al.}~\cite{Jouguet2011} adds the required constraints for multidimensional reconciliation on top of Stewart method: the built random transformation must map Bob's quantum channel data $\vec{b}$ to the input of privacy amplification $\vec{u}$, and the description of the transformation (disclosed on the classical channel) must not reveal additional data to Eve.

We use the following notations:
\begin{itemize}
    \item The algorithm takes as input normalised vectors $\vec{b}$ and $\vec{u}$ in $\mathbb{R}^d$.
    \item For $i$ from $d$ to $1$, the vector $\lsub{i}{n} \in \mathbb{R}^i$ describes a (modified) Householder transformation $\lsu{i}{\tilde{H}}$ in the subspace composed of the last i coordinates of $\mathbb{R}^d$. $\lsu{i}{H}$ is its extension to the whole space $\mathbb{R}^d$.
    \item $\lsu{i}{Q}$ = $\lsu{d}{H} \circ \lsu{d-1}{H} \circ \ldots \lsu{i}{H}$ is the composition of the transformations from $i$ to $d$.
    \item $\lsub{d}{u} = \vec{u}$, and for $i < d$, $\lsub{i}{u}$ is the last i coordinates of $\lsu{i}{Q}(\vec{u})$. ($\lsub{i}{u}$ can be computed only once the $\lsu{i}{H}$ from previous iterations are known.)
    \item We also note $\prescript{}{i}{\vec{e}}_1$ the first vector of the canonical basis of $\mathbb{R}^i$ and $\lsub{i}{b}$ the last $i$ coordinates of $\vec{b}$..
\end{itemize}

The algorithm is then the following. For $i$ from $d$ to $2$:
\begin{enumerate}
    \item Build a random unit vector $\lsub{i}{g} \in \mathbb{R}^i$ such that $\lsub{i}{g} \cdot \lsub{i}{u} = \lsub{i}{e_1} \cdot \lsub{i}{b} = b_{d-i+1}$. A detailed method is given at the end of this section.
    \item Find $\tilde{H}_i$ the Householder transformation that maps $\lsub{i}{g}$ to $\prescript{}{i}{\vec{e}}_1$: $\displaystyle \lsub{i}{n} = \frac{\lsub{i}{g}-\prescript{}{i}{\vec{e}}_1}{\|\lsub{i}{g}-\prescript{}{i}{\vec{e}}_1\|}$.
\end{enumerate}

For $i=1$, pick the transformation that multiplies by $\text{sign}(\lsub{1}{u}/b_d)$: $\lsub{1}{n} \in \{0, 1\}$.

In the multidimensional reconciliation context, Bob disclose the vectors $\lsub{i}{n}$ to Alice, who iteratively apply the $d$ transformations on here quantum channel data $\vec{a}$ to compute her noisy version of $\vec{u}$.

\textbf{Building $\lsub{i}{g}$}
\begin{enumerate}
    \item Generate a random vector $\lsub{i}{h} \in \mathbb{R}^i$ by drawing $i$ coefficients from a standard normal distribution.
    \item Find $\lsu{i}{\alpha} \in \mathbb{R}$ such that $\lsub{i}{u} \cdot (\lsub{i}{h} + \lsu{i}{\alpha}   \lsub{i}{u}) = b_{d-i+1} \|\lsub{i}{h} + \lsu{i}{\alpha} \lsub{i}{u}\|$.

          By squaring it, the equation can be written as
          \begin{align*}
               & \lsu{i}{\alpha}^2 (Y^2 - b_{d-i+1}^2 Y) & +   \\
               & \lsu{i}{\alpha}(2XY - 2b_{d-i+1}^2X)    & +   \\
               & (X^2-b^2Z)                              & = 0
          \end{align*}

          where:
          \begin{itemize}
              \item \( X = \lsub{i}{u} \cdot \lsub{i}{h} = \lsub{i}{h} \cdot \lsub{i}{u}, \)
              \item \( Y = \| \lsub{i}{u} \|^2, \)
              \item \( Z = \| \lsub{i}{h} \|^2. \)
          \end{itemize}

          This quadratic equation can have two solutions. However, since it is obtained by squaring the original constraint, the sign information is lost. Hence, only one of the two solutions satisfies the original equation.
          The valid solution is the one such that
          \[
              \operatorname{sign}\!\big(\lsub{i}{u}\cdot(\lsub{i}{h}+\alpha\,\lsub{i}{u})\big) = \operatorname{sign}(b_{d-i+1}),
          \]

          Therefore, the appropriate root must be chosen according to the sign of \( b_{d-i+1} \). In practice,
          \[
              \lsu{i}{\alpha} =
              \begin{cases}
                  \dfrac{-B + \sqrt{\Delta}}{2A}, & \text{if } b_{d-i+1} \geq 0, \\
                  \dfrac{-B - \sqrt{\Delta}}{2A}, & \text{otherwise},
              \end{cases}
          \]

          where:
          \begin{itemize}
              \item \( A = Y^2 - b_{d-i+1}^2 Y, \)
              \item \( B = 2XY - 2b_{d-i+1}^2X, \)
              \item \( C = X^2-b^2Z, \)
              \item \( \Delta = B^2-4AC. \)
          \end{itemize}

\end{enumerate}

$\lsub{i}{g}$ is then defined as $\displaystyle \lsub{i}{g} = \frac{\lsub{i}{h} + \lsu{i}{\alpha} \lsub{i}{u}}{\|\lsub{i}{h} + \lsu{i}{\alpha} \lsub{i}{u}\|}$.

\section{Virtual channel mutual information}
\label{appendix:virtual_theory}

We give in this appendix more details on the mathematical description of the virtual channel. We work directly in $\mathbb{R}^d$ and simplify the notation dropping $d$ and $i$: ${}_d\vec{x}_i$ is noted $\vec{x}$.

For direct reconciliation, $\vec{u}_\text{direct}$ and $\R_\text{direct}$ are generated by Alice. Applying $\R_\text{direct}$ to equation (\ref{eq:channel}) gives for the virtual channel output
\begin{align}
    \vec{v}_\text{direct} = \R_\text{direct}(\vec{b}) & = \R_\text{direct}(t \vec{a} + \vec{z})                            \\
                                                      & = \frac{t \|\vec{a}\|}{\sqrt{d}} \vec{u}_\text{direct} +\R\vec{z}.
\end{align}
For reverse reconciliation applying $\R$ to equation (\ref{eq:reverse_channel}) gives
\begin{equation}
    \label{eq:virtual_channel}
    \vec{v} = \frac{t\sigma_a^2} {T\sigma_a^2+\sigma_z^2} \frac{\|\vec{b}\|}{\sqrt{d}} \vec{u} +\R\vec{w}.
\end{equation}
Our goal is to show that these equations can be interpreted as $d$ independent BIAWGN channels. We focus here on the reverse scenario.

$\vec{b}$ and $\vec{w}$ are independent isotropic Gaussians and $\vec{u}$ is uniform on $\{1, -1\}^d$ and independent of $(\vec{b}, \vec{w})$. $\R$ is constructed as in \ref{subsec:QR}: $\R = H\circ Q$ where $Q$ is a Haar-random orthogonal matrix independent of $(\vec{b}, \vec{w}, \vec{u}$), and $H$ is the Householder transformation that maps $Q\vec{b}$ to $ \frac{\|\vec{b}\|}{\sqrt{d}}\vec{u}$.

The key ingredients are the invariance of the laws of $\vec{b}$ and $\vec{w}$ under every orthogonal map (isotropy), and the invariance of Haar probability on $\mathcal{O}(d)$ under multiplication by a fixed orthogonal map.

By isotropy, replacing $\vec{b}$ by $Q\vec{b}$ preserves its law and its independence from $Q$, $\vec{u}$ and $\vec{w}$.

Conditioning on $Q\vec{b}$ and $\vec{u}$ makes $H$ fixed while $Q$ remains Haar distributed. Haar invariance therefore leaves $\R$ Haar distributed conditionally on $(Q\vec{b}, \vec{u})$, with a conditional law independent on their values. $\R$ is thus Haar distributed and independent of these variables.

Moreover, isotropy of $\vec{w}$ preserves its law when it is rotated by $\R$.

We can therefore write the factorization
\begin{align*}
     & \mathcal{L}\bigl(\|\vec{b}\|,\R,\vec{u},\R\vec{w}\bigr) =
    \\
     & \mathcal{L}(\|\vec{b}\|)
    \otimes \mathrm{Haar}_{\mathcal{O}(d)} \otimes \mathrm{Unif}(\{-1,1\}^d)
    \otimes \mathcal{N}(0,\sigma_w^2\mathbb{I}_d),
\end{align*}
where $\mathcal{L}(X)$ denotes the probability law of $X$.

Thus, after revealing both $\|\vec{b}\|$ and $\R$, $\vec{u}$ remains uniform, $\R\vec{w}$ remains an isotropic Gaussian vector, and the two remain independent. Equation (\ref{eq:virtual_channel}) consequently corresponds to $d$ independent BIAWGN channels.

The corresponding log-likelihood ratio is
$$\mathrm{LLR} = \frac{2t\|\vec{b}\|}{\sqrt{d}}\frac{v_i}{\sigma_z^2},$$
and the signal-to-noise ratio is
$$
    \text{SNR}_{d, \|\vec{b}\|} = \frac{\|\vec{b}\|^2}{T\sigma_a^2 + \sigma_z^2} \frac{\text{SNR}}{d}.
$$

Denoting by $I_{\text{BIAWGN}}(\zeta)$ the mutual information of the BIAWGN channel with signal-to-noise ratio $\zeta$, the mutual information for fixed public information is $I_{\mathrm{BIAWGN}}(\text{SNR}_{d, \|\vec{b}\|})$.

Now, considering a codeword that spans many independent blocks of size $d$, the relevant quantity is the ergodic mutual information, averaging over the SNR. Because $\|\vec{b}\|^2 / (T\sigma_a^2 + \sigma_z^2)$ is $\chi^2_d$ distributed, the ergodic mutual information of the virtual channel is

$$
    I_d = \mathbb{E}_{\lambda \sim \chi_{d}^2}\left(I_{\text{BIAWGN}}\left(\frac{\lambda}{d}\text{SNR}\right)\right)
$$

Because the direct and reverse reconciliation channels have the same structure, it can be shown similarly that the same expression is valid for the direct reconciliation virtual channel.

\clearpage

\printbibliography

\end{document}